\documentclass[namedreferences]{SolarPhysics}
\usepackage[optionalrh]{spr-sola-addons} 
\usepackage{graphicx}
\usepackage{color}           
\usepackage{url}             

\newcommand{\adv}{    {\it Adv. Space Res.}}

\newcommand{\aap}{    {\it Astron. Astrophys.}}
\newcommand{\aaps}{   {\it Astron. Astrophys. Suppl.}}

\newcommand{\aj}{     {\it Astronom. J.}}
\newcommand{\apj}{    {\it Astrophys. J.}}
\newcommand{\apjl}{   {\it Astrophys. J. Lett.}}

\newcommand{\grl}{    {\it Geophys. Res. Lett.}}

\newcommand{\jgr}{    {\it J. Geophys. Res.}}

\newcommand{\pasj}{   {\it Pub. Astron. Soc. Japan}}

\newcommand{\solphys}{{\it Solar Phys.}}

\newcommand{\ssr}{    {\it Space Sci. Rev.}}

\begin{document}

\begin{article}

\begin{opening}

\title{Coronal Shock Waves, EUV waves, and Their Relation to CMEs. I. Reconciliation of ``EIT waves'', Type II Radio Bursts, and Leading
Edges of CMEs}

\author{V.V.~\surname{Grechnev}$^{1}$\sep
        A.M.~\surname{Uralov}$^{1}$\sep
        I.M.~\surname{Chertok}$^{2}$\sep
        I.V.~\surname{Kuzmenko}$^{3}$\sep
        A.N.~\surname{Afanasyev}$^{1}$\sep
        N.S.~\surname{Meshalkina}$^{1}$\sep
        S.S.~\surname{Kalashnikov}$^{1}$\sep
        Y.~\surname{Kubo}$^{4}$
        }

\runningauthor{Grechnev et al.}
 \runningtitle{``EIT waves'' and Type II Radio Bursts}

\institute{${}^{1}$Institute of Solar-Terrestrial Physics SB RAS,
Lermontov St.\ 126A, Irkutsk 664033, Russia email: \url{grechnev@iszf.irk.ru} \\
    $^{2}$  Pushkov Institute of Terrestrial Magnetism,
            Ionosphere and Radio Wave Propagation (IZMIRAN), Troitsk, Moscow
            Region, 142190 Russia email: \url{ichertok@izmiran.ru}\\
    $^{3}$ Ussuriysk Astrophysical Observatory, Solnechnaya St. 21,
            Primorsky Krai, Gornotaezhnoe 692533, Russia
                  email: \url{kuzmenko_irina@mail.ru}\\
    $^{4}$ National Institute of Information and Communications
            Technology, Tokyo, Japan email: \url{kubo@nict.go.jp}
}

\begin{abstract}
We show examples of excitation of coronal waves by flare-related
abrupt eruptions of magnetic rope structures. The waves presumably
rapidly steepened into shocks and freely propagated afterwards
like decelerating blast waves that showed up as Moreton waves and
EUV waves. We propose a simple quantitative description for such
shock waves to reconcile their observed propagation with drift
rates of metric type II bursts and kinematics of leading edges of
coronal mass ejections (CMEs). Taking account of different plasma
density falloffs for propagation of a wave up and along the solar
surface, we demonstrate a close correspondence between drift rates
of type II bursts and speeds of EUV waves, Moreton waves, and CMEs
observed in a few known events.

\end{abstract}


\end{opening}


\section{Introduction}

Some solar eruptions are accompanied by large-scale wave-like
disturbances visible in various spectral ranges. Moreton waves
\cite{Moreton1960} observed in the H$\alpha$ line have been
initially proposed by \inlinecite{Uchida1968} to be a
chromospheric trail of a coronal fast-mode MHD wave. Observations
of the low corona in extreme ultraviolet (EUV) with EIT
\cite{Delab1995} on SOHO revealed large-scale wave-like
disturbances visible as fronts of enhanced (but still low)
brightness, either quasi-stationary or propagating over large
distances up to the whole disk along the solar surface or
expanding above the limb. These transients called ``EIT waves''
(or ``EUV waves'') are registered by a number of EUV telescopes
--- EIT, TRACE, STEREO/EUVI, SDO/AIA, \textit{etc.} (see
\opencite{Warmuth2007}; \opencite{WillsDavey2009};
\opencite{Gallagher2010} for a review).

Analysis of observations and interpretation of such phenomena meet
problems. Glaring flare emission hampers detection of faint EUV
waves. Many data are limited by 12-min imaging rate of EIT. Faster
TRACE observations \cite{Handy1999} have a small field of view. It
is often difficult to reliably identify and trace a moving feature
of interest. To overcome these difficulties, special methods are
employed, but they might contribute artifacts. Multi-instrument
analyses sometimes encounter timing problems. Thus, some results
used in interpretation and modeling might not be completely
adequate to the observed phenomena.

The nature of EUV waves has been debated starting from their
discovery \cite{Thompson1998}. The most popular interpretations of
a near-surface EUV wave are \textit{i})~traces of an MHD fast-mode
wave in the lower corona (\textit{e.g.}, \opencite{Thompson1999};
\citeauthor{Warmuth2001} \citeyear{Warmuth2001,Warmuth2004b};
\opencite{KhanAurass2002}; \opencite{HudsonWarmuth2004}) and
\textit{ii})~plasma compression in bases of coronal loops in their
successive stretching by an expanding CME (\textit{e.g.},
\opencite{Delannee2000}). A numerical 2D MHD simulation of a
magnetic flux rope eruption \cite{Chen2005} revealed both the
(\textit{i}) and (\textit{ii}) disturbances.
\inlinecite{Schmidt2010} presented the first 3D MHD modeling of an
``EIT wave'' as a disturbance produced by an eruption-driven shock
wave. A fast-mode wave detected in the simulation corresponded to
an EUV wave observed in the modeled event including reflection
from a coronal hole in support of the wave hypothesis.

None of existing models describes all properties of EUV waves.
Most likely, this is because the multitude of transients observed
as EUV waves actually correspond to different phenomena. This
conjecture is supported by a variety of morphologic and dynamic
characteristics of observed EUV waves. For example, 1)~their
velocities estimated from observations of some events exceeded the
coronal fast-mode speed, whereas they were lower in other events
\cite{WillsDavey2007,Warmuth2010}; 2)~the wave front can be either
diffuse or sharp; 3)~kinematics of an EUV wave can be incompatible
with the fast-mode MHD wave model \cite{Zhukov2009}. On the other
hand, such properties of EUV waves as deceleration, decay, and
broadening the disturbance (\citeauthor{Warmuth2001}
\citeyear{Warmuth2001,Warmuth2004a,Warmuth2004b};
\opencite{Long2008}; \opencite{Veronig2010}), bypassing regions of
an increased Alfv{\' e}n velocity --- coronal holes and active
regions \cite{Thompson1999}, possible reflections
\cite{VeronigTemmerVrsnak2008,Gopalswamy2009} appear to correspond
to the hypothesis of a coronal MHD wave. Considerations of
wave-like transients sometimes observed in EUV to expand above the
limb also suggest that different phenomena might be involved (see,
\textit{e.g.}, \opencite{Zhukov2004};
\citeauthor{Grechnev2006fil},
\citeyear{Grechnev2006fil,Grechnev2008};
\opencite{Pohjolainen2008}; \opencite{Meshalkina2009};
\opencite{Chertok2009}; \opencite{Cohen2009}). In paper~III
\cite{Grechnev2010III} we consider an event with a two-component
``EIT wave''; a propagating component matched properties of a
coronal shock wave, while a stationary component was presumably
associated with a stretched CME structure.

If some EUV waves are really due to coronal shock waves, then
their correspondence with signatures of shocks in higher corona is
expected. Type II radio bursts are considered as manifestations of
shock waves propagating upwards in the corona (\textit{e.g.},
\opencite{VrsnakCliver2008}). \inlinecite{Klassen2000} concluded
that almost all metric type II bursts were accompanied by EUV
waves but stated no correlation between their speeds; on average,
the speeds of EUV waves were three times less than estimates from
drift rates of type II bursts. \inlinecite{Biesecker2002} found
that many EUV waves were not associated with type II bursts.

\inlinecite{Sheeley2000} proposed that kinks or deflections of
coronal rays at the flanks and rear ends of fast CMEs could be
signatures of shock waves. An important common property of these
flank/rear kinks was their deceleration, while the authors did not
reveal deceleration of CME leading edges\footnote{The SOHO LASCO
CME Catalog at \url{http://cdaw.gsfc.nasa.gov/CME_list/}
(afterwards `the CME Catalog'; \opencite{Yashiro2004}) shows
pronounced decelerations for three of the four events considered
by the authors.}. A usually considered scenario for the formation
of shock waves, which might show up in LASCO images, seems to be
incompatible with wave signatures on the solar surface. In this
scenario, when the CME speed exceeds the local fast-mode speed at
heliocentric distances $\gsim 1.5R_{\odot}$, a bow shock forms
continuously pressed by a fast CME. However, a bow shock followed
by a Mach cone can only be formed by a supersonic body of a fixed
size, whereas CMEs expand omnidirectionally (with respect to their
center). Hence, the conic bow shock geometry appears to be
unlikely for wide CMEs. Since neither mechanisms nor heights of
the shock formation have been established, possible association
between surface EUV waves and CME components cannot be excluded.
Indeed, \inlinecite{Veronig2010} found the upper part of an
expanding EUV dome to coincide with a white-light CME, while the
lower skirt of the dome was a surface EUV wave.

For these reasons it is difficult to expect that all observed
properties of EUV waves could be explained by a single mechanism.
Based on this assumption, our three companion papers are focused
on those EUV waves, which are most likely associated with coronal
shock waves. We address a few events, all of which were previously
studied. Even for a rather uniform subset of phenomena probably
corresponding to similar processes, seemingly contradictions
between some observational facts might occur. We endeavor to
reconcile such discrepancies.

One of challenging issues is reconciliation of EUV wave velocities
and drift rates of type II bursts. This is a subject of the
present paper I, the first one of the three companion papers.
Assuming the shock wave nature of EUV waves under consideration,
we try to settle disagreement between different studies. In this
paper we use a simplest approximation of a self-similar shock wave
that is convenient in comparisons with observations, which often
show self-similarity of the wave front expansion. However, this
approximation corresponds to the strong shock wave limit and
cannot apply to all stages of events. A weaker shock seems to be
more realistic, but its calculations are more complex. Propagation
of a weak shock along the solar surface is analytically modeled in
Paper~II \cite{Afanasyev2010}. Paper~III analyzes propagation of a
probable shock wave in the 17 January 2010 event using both strong
and weak shock approximations.

\section{Methodical Issues}

Excitation of shock waves in eruptive events seems to be
undoubted, but their sources have not been established. Three
possible exciters of shock waves are considered: \textit{i})~a
pressure pulse produced by a flare, \textit{ii})~a super-Alfv{\'
e}nic piston, and \textit{iii})~an impulsive piston. In case
(\textit{ii}), a bow shock continuously driven by a supersonic CME
is expected to appear with kinematics determined by the driver. In
case (\textit{i}) or (\textit{iii}), the shock wave expelled by an
impulsive driver propagates afterwards freely like a decelerating
blast wave (\textit{cf.} \opencite{PomoellVainioKissmann2008}).
Observations suggest that shock waves excited by impulsive drivers
and freely propagating in the low corona do exist. This is
indicated by deceleration of Moreton/EUV waves (\textit{e.g.},
\citeauthor{Warmuth2001} \citeyear{Warmuth2001,Warmuth2004a}),
differences between propagation directions of the wave and a
possible driver \cite{Hudson2003}, very early appearance of type
II bursts. We accept this possibility as a working hypothesis and
use a description of propagation of a blast shock wave. It is
possible to calculate it analytically for two limits. One limit is
a strong self-similar wave, whose length along a propagation
direction is comparable with the curvature radius of the wave
front. The opposite limit is a weak shock wave, whose length is
much less than the curvature radius of the front and a typical
size of inhomogeneities in the medium. \inlinecite{Grechnev2008}
found that a formal usage of expressions for propagation of a
strong self-similar shock wave excited by a point-like explosion
in a gas allowed to fit the speeds and positions of a Moreton wave
as well as an EUV wave at the initial stage of the motion.

\subsection{Self-Similar Shock Wave Approximation}
 \label{S-approximation}

\inlinecite{Grechnev2008} used a simple model to describe
propagation of such a blast-like wave in plasma with a radial
power-law (PL) density falloff $\delta$ from an eruption center,
$n = n_0(x/h_0)^{-\delta}$ with $x$ being the distance and $n_0$
the density at a distance of $h_0$. We use $h_0 \approx 100$ Mm,
close to the scale height. (\textit{Our notations are different
from those used in the papers listed above}). Self-similarity of a
shock wave is ensured by a presumable large pressure excess inside
the volume confined by the shock front over a non-disturbed
medium. Propagation of a shock wave in the self-similar
approximation is determined by plasma density distribution being
almost insensitive to the magnetic fields. Such a wave decelerates
when $\delta < 3$ due to a growing mass of swept-up material.
Propagation of a strong shock vs. time $t$ in plasma with a PL
density model is described by an expression
\begin{equation}
x(t) \propto t^{2/(5-\delta)}
 \label{E-wave_prop}
\end{equation}
The approximation has a singularity at $x \to 0$ (here also the
wave velocity $v \to \infty$); however, wave signatures are not
observed at small distances. The approximation becomes inaccurate
at large distances, being not limited from below by the fast-mode
speed. A wave traveling along the solar surface weakens at large
distances and propagates, in the first approximation, in a
flat-layered atmosphere. Expression (\ref{E-wave_prop}) describes
strong spherical shock waves, which seem to be unrealistic in
solar conditions, but its usage within some range of distances can
be justified. 1)~An enhanced plasma density above an active region
falls off both vertically and horizontally. A power-law
description of the falloff seems to be acceptable. 2)~The
self-similar solution of a strong wave satisfactorily describes
damping of a gas-dynamic shock wave up to Mach numbers $M \approx
2$, when the wave is neither strong nor weak. 3)~Applicability of
gas-dynamic self-similar solutions to MHD blast shock waves is not
obvious, because account of the magnetic field seems to be
necessary. We note the following. With $M \gg 2$, the gas pressure
behind the shock front exceeds the magnetic pressure, even if
$\beta = C_\mathrm{S}^2/V_{\mathrm{A}}^2 \ll 1$ in non-disturbed
plasma before the shock front; here $M$ is a ratio of the shock
speed to the fast-mode speed before the front, $C_\mathrm{S}$ and
$V_{\mathrm{A}}$ are the sound and Alfv{\' e}n speeds. That is,
the plasma flow behind the shock front has a gas-dynamic
character. The role of magnetic fields is also not crucial for
medium-intensity shocks ($M \geq 2$), which also strongly heat
plasma, thus significantly increasing its pressure. For example,
with $M \approx 2$, the plasma pressure behind the front of a wave
perpendicular to the magnetic field is equal to the magnetic
pressure before the front, even if $\beta \ll 1$ there. For a
switch-on shock wave running along the magnetic field this occurs
with a Mach number $M \approx 1.5$.

Expression (\ref{E-wave_prop}) was obtained under an assumption
that the $\delta$ index was independent of the wave propagation
direction $\vartheta$. We will formally use this expression also
when $\delta = \delta(\vartheta)$, if variations of $\delta$ are
small with the change of the direction, \textit{i.e.}, $\pi
d\delta \ll d\vartheta$. Note that in a limit of a weak, short
shock wave, its propagation is determined by a local value of
$\delta$ even if this condition is not satisfied. The above
considerations lead to a heuristic conclusion about a possibility
to use expression (\ref{E-wave_prop}) for approximate estimates of
kinematic characteristics of shock waves of intermediate intensity
propagating in medium with $\delta = \delta(\vartheta)$.

It is useful to compare the power-law coronal density model with
other popular models. The \inlinecite{Newkirk1961} model ($n_e =
4.2 \times 10^4 \times 10^{4.32/r}$, $r$ is the heliocentric
distance expressed in solar radii) describes the radial plasma
density distribution in a coronal streamer. The Saito model
\cite{Saito1970} describes the density distribution above the
quiet Sun depending on the latitude $\phi$
\begin{equation}
{\frac {n_e(r, \phi)} {10^8}} =
 {\frac{3.09} {r^{16}}}(1-0.5\sin{\phi}) +
 {\frac{1.58} {r^{6}}} (1-0.95\sin{\phi}) +
 {\frac{0.0251} {r^{2.5}}}(1-{\sin^{0.5}{\phi}})
  \label{E-Saito_model}
\end{equation}

  \begin{figure} 
  \centerline{\includegraphics[width=\textwidth]
   {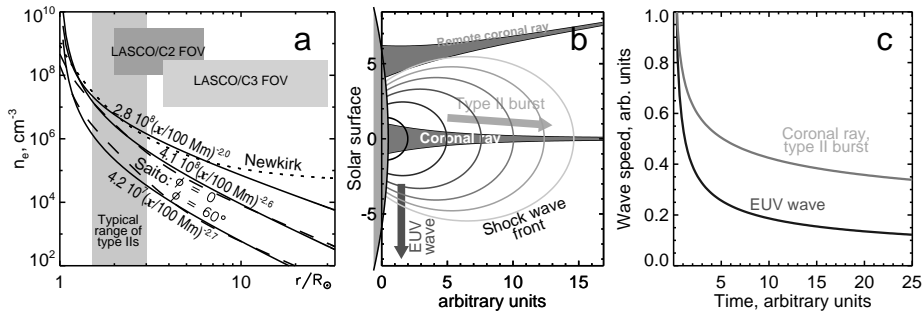}
  }
  \caption{a)~Coronal density models of Newkirk (dotted) and Saito (dashed, for
latitudes $\phi = 0^{\circ}$ and $\phi = 60^{\circ}$). The solid
lines represent the power-law model with different parameters to
fit the Newkirk model at $r < 9R_{\odot}$ and the Saito model
within the LASCO/C2 \& C3 fields of view shown with shading. The
vertical shaded region indicates the typical range of type II
bursts. b)~A cartoon illustrating the relation between a type II
burst and an EUV wave. A narrowband type II emission is generated
by a shock front propagating in a distinct extended narrow
structure like a coronal ray. A near-surface EUV wave runs slower.
(c)~Speed\,--\,time plots of the EUV wave and the type II emission
site.
    }
  \label{F-density_models_cartoon1}
  \end{figure}

Figure~\ref{F-density_models_cartoon1}a presents the Saito model
for $\phi = 0^{\circ}, 60^{\circ}$, and the Newkirk model. The PL
model can be adjusted to any of these models by varying its
parameters. The $n_0, h_0$ parameters are redundant; we have split
them to clarify their physical meaning. The $x$ variable in the
radial direction is $x \approx (r-1)R_{\odot}$. The PL model with
$\delta = 2$, $n_0 = 2.8 \times 10^8$ cm$^{-3}$ agrees within $\pm
30\%$ with the Newkirk model at $r = (1.2-9)R_\odot$, which is
important in anaylses of type II bursts. The parameters of the PL
model can be adjusted to the Saito model for various $\phi$ as
well, which is important in analyses of CMEs. A single PL model
with a direction-dependent $\delta$ provides a convenient
alternative to complex involvement of various density models and
allows one to account for individual properties of active regions
as well as highly disturbed conditions just before the onset of a
wave.

\subsection{Shock Waves and Type II Bursts}
 \label{S-shock_type_II}

Assuming $\delta$ to depend on a propagation direction, we get an
approximation for a shock of an intermediate intensity propagating
in an anisotropic medium
(Figure~\ref{F-density_models_cartoon1}b). A quasi-isotropic shock
wave propagating in homogeneous corona can only cause drifting
continuum radio emission. A strong narrowband harmonic type II
burst can appear if the shock front passes along a lengthy
structure like a coronal streamer (\textit{e.g.},
\opencite{Reiner2003}; \opencite{MancusoRaymond2004}). The
cumulation effect increases the density jump in vicinities of the
streamer's current sheet and intensifies radio emission
\cite{Uchida1974}. The situation resembles a flare process running
along a coronal ray \cite{Uralova1994}. The difference between the
horizontal and vertical directions explains a relation between the
speeds of an EUV wave and a type II burst
(Figure~\ref{F-density_models_cartoon1}c). The upwards speed is
higher than the surface one, because the wave center rises.

Even if a shock wave appears at a zero height, its front rapidly
becomes convex and tilted towards the solar surface
(Figure~\ref{F-density_models_cartoon1}b). This front shape has
been actually observed by \inlinecite{Hudson2003} and is
consistent with observations discussed by
\inlinecite{Warmuth2004b}. If a convex shock front encounters a
remote coronal ray, then the intersection site bifurcates, and its
parts move along the ray in opposite directions (\textit{cf.}
\opencite{MancusoAbbo2004}). The contact corresponds to an
infinite drift rate followed by bidirectional drifts to resemble a
direct bracket ``('' in a dynamic spectrum. Note that dynamic
spectra present a combination of emissions originating at
different sites, so that the intensities are summed.

In Section~\ref{S-observations} we reconcile kinematics of ``EIT
waves'' and drift rates of corresponding type II bursts in terms
of our approach based on the self-similar shock approximation
(Section~\ref{S-approximation}). We apply power-law curves with
the same onset time to both spectral domains (hereafter `shock-PL
fit'). The density falloff index in a streamer determining the
drift rate of a type II burst is expected to be $\delta \sim 2$.
Real lateral density falloffs in a streamer should be steeper than
along its axis. Otherwise, streamers would not be visible in
homogeneous corona. Thus, real shock fronts should be oblate at
medium distances. The density falloff for an EUV wave escaping
from an active region can be $0 < \delta <3$. For possible wave
signatures in CMEs, the density falloff index is expected to be
close to the Saito model (\ref{E-Saito_model}), i.e., $\delta \sim
2.6$ at moderate latitudes and steeper at higher latitudes.

\subsection{Fit of Presumable Wave Signatures}

We fit the drift rate of a type II burst manually. The onset time
$t_0$ of a wave can be approximately estimated from observations.
We take parameters of the plasma density model $n_0 = 5.5 \times
10^8 $~cm$^{-3}$, $h_0 = 100$~Mm, and $\delta$ according to the
considerations in the preceding Section. One more input parameter
is a reference frequency $f_\mathrm{obs}$ of a band with a
harmonic number $N_f$ (usually 1 or 2) actually observed in a
dynamic spectrum at a time $t_1$. The corresponding plasma number
density is $n_1 = [f_\mathrm{obs}(t_1)/N_f/(0.9 \times 10^4
)]^{2}$, and the height is $x_1 = h_0\,(n_0/n_1)^{1/\delta}$. Then
the height\,--\,time plot of the shock tracer is calculated as
$x(t) = x_1\,[(t - t_0)/(t_1 - t_0)]^{2/(5-\delta)}$, the
corresponding density variation as $ n(t) =
n_0\,[x(t)/h_0]^{-\delta}$, and the outline of both bands of the
type II burst as $f(t) = [1, 2] \times 0.9 \times 10^4
[n(t)]^{1/2}$. By adjusting $\delta$ and $t_0$ in sequential
attempts, we endeavor to approach a best outline of the bands.
Uncertainties provided by the routine typically do not exceed
1~min for $t_0$ and 0.2 for $\delta$. The spectrum can be
coordinated with measured heights by adjusting $n_0$, as usually
done.

To fit presumable traces of shocks in coronagraph images (in this
paper we use measurements from the CME Catalog), we employ two
ways. The first way is a manual fit similar to the routine
outlined in the preceding paragraph. Input parameters are starting
estimates of $\delta$ and $t_0$, the heliocentric distances of the
wave origin $r_0$ and the wave front $r_1$ measured at a time
$t_1$. The initial approximation of the height\,--\,time plot is
$r(t)=(r_1-r_0)\left[(t-t_0)/(t_1-t_0)\right]^{2/(5-\delta)} +
r_0$. Then sequential attempts are made to approach a best fit.
The second way employs a log\,--\,log height\,--\,time plot, which
is a straight line for a power law. We use a second-order fit and
iteratively vary $\delta$ and $t_0$ to minimize the second-order
term. One should be aware of the fact that the major but unknown
uncertainties can be due to identification of the wave front in
coronagraph images.

\section{Observations}
 \label{S-observations}

\subsection{Event 1: 13 July 2004}
 \label{S-event1}

\begin{figure} 
  \centerline{\includegraphics[width=\textwidth]
   {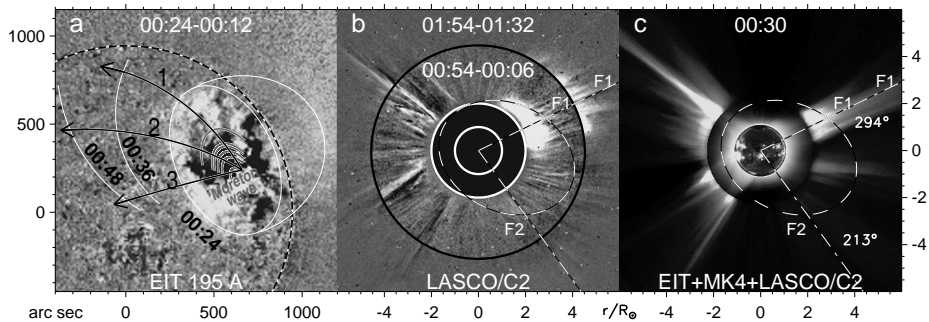}
  }
  \caption{The 13 July 2004 event. a)~The Moreton wave (gray) and EUV
wave (white) fronts superposed on the EIT 195~\AA\ difference
image. The black arcs  1, 2, and 3 trace the directions of
measurements. The dotted circle denotes the limb. (b)~The leading
part of the CME observed in two LASCO/C2 difference images at
00:54 and 01:54 separated with the black circle. c)~The
non-disturbed corona in a combined image composed from
non-subtracted EIT 195~\AA, Mark4, and C2 images. The dashed ovals
in (b) and (c) outline the CME in the 00:54 image. The dash-dotted
straight lines in (b) and (c) mark the position angles at which
the measurements listed in the CME Catalog were carried out for
the fastest features F1 and F2. The axes show the distances from
the solar disk center in arc seconds (a) and in solar radii (b and
c).
    }
  \label{F-20040713-eit_lasco}
  \end{figure}

This event (Figure~\ref{F-20040713-eit_lasco}) was associated with
an eruptive M6.7 flare (00:09--00:23, \textit{all times hereafter
are UT}) in active region 10646 (N13\thinspace W46) and a CME
observed with SOHO/LASCO \cite{Brueckner1995} after 00:54. Two
parts of the CME (Figure~\ref{F-20040713-eit_lasco}b,c) are listed
in the CME Catalog as two CMEs measured at position angles (PA)
of $294^{\circ}$ and $213^{\circ}$. A type II burst was recorded
in three observatories. The three estimates of the shock speed
progressively decreased in time.

\inlinecite{Grechnev2008} revealed signatures of a probable blast
wave as an H$\alpha$ Moreton wave (gray in
Figure~\ref{F-20040713-eit_lasco}a) and an EUV wave (white). Both
disturbances were kinematically close to each other and to a
kinematical curve expected for a lower trail of a decelerating
coronal blast wave. Its exciter was not discussed. The authors
proposed that the decreasing estimates of the shock speed
reflected deceleration of a single shock wave, but they did not
consider the type II burst.

\inlinecite{Pohjolainen2008} [afterwards PHS] analyzed the type II
burst in this event, but could not reconcile the overall drift
with propagation of a single shock wave. The authors proposed that
two shock waves were excited, one by a flare blast, and the second
one by an expanding loop, a part of a CME.

We carried out an additional analysis of this event, measured
kinematics of an eruptive system in order to find out a probable
origin of a shock wave(s), and to reconcile its (their)
propagation with the EUV/Moreton waves and the CME.

\subsubsection{Eruptive System}
 \label{S-eruptive_system}
\begin{figure} 
  \centerline{\includegraphics[width=\textwidth]
   {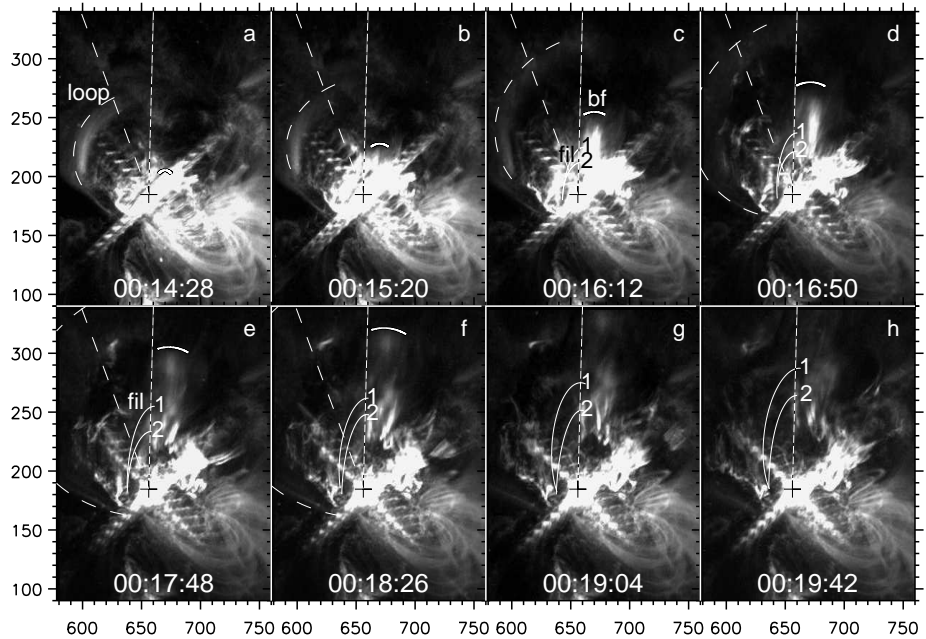}
  }
  \caption{The eruptive system in the 13 July 2004 event (TRACE,
173~\AA). The cross marks the initial position of the filament.
Oval arcs outline the eruptive loop (dashed), filaments (solid),
and bright feature (short arc). The broken lines denote the
expansion directions of the loop (dashed) and the filament
(dotted). Axes show arc seconds from the solar disk center.
  }
  \label{F-20040713-eruptive_system}
  \end{figure}

Figure~\ref{F-20040713-eruptive_system} shows the eruptive system:
a leading bright feature `bf', two filament segments `1' and `2',
and several eruptive loops, of which one (`loop') was conspicuous
and thus will be discussed henceforth. Long exposure times
(33\,--\,46~s) caused a blurring of fast features, \textit{e.g.},
a jetlike appearance of the bright feature.

To measure expansion of a feature, we outline it with an oval arc
that allows us to trace it, even if its leading edge is difficult
to detect. The obtained distance\,--\,time plot is used as a
starting estimate. Then we choose a regular function to match the
distance\,--\,time plot and estimate its parameters. Using the
analytic fit, we calculate expected distance\,--\,time points,
compare them with observations, and improve the fit. All
kinematical plots are calculated by means of integration or
differentiation of the analytic fit rather than measurements. Our
ultimate criterion is to reproduce the motion of an analyzed
feature. Observational limitations do not allow us to reveal a
detailed time profile of acceleration, and we describe it instead
with a smooth bell-like function. We use a Gaussian time profile
(see \opencite{WangZhangShen2009}). Then acceleration $a$ is
\begin{eqnarray}
 a = \left( v_1-v_0 \right) \exp{\{-{[(t-t_0)/\tau_{\mathrm{acc}}]^2}/2 \}} /
  (\sqrt{2\pi}\tau_{\mathrm{acc}})
 \label{E-acceleration}
\end{eqnarray}
Here $\tau_{\mathrm{acc}}\sqrt{8\ln{2}}$ is a full width at
half-maximum of the acceleration time profile, which is centered
at the $t_0$ time; $v_0$ and $v_1$ are velocities at the onset and
end of the acceleration stage. In cases of a more complex
kinematics, we use a combination of Gaussians and adjust their
parameters manually.

Uncertainties are mainly determined by difficulties to identify
and trace a feature in question. For this purpose we use both
non-subtracted and difference images processed in various ways and
improve results in sequential attempts. The worst traceable
feature in this event is the eruptive filament, which appears as a
semitransparent dark feature, whose segments 1 and 2 are faintly
visible in Figure~\ref{F-20040713-eruptive_system}. We therefore
present the measurements of the filament in more detail.

  \begin{figure}    
   \centerline{
               \includegraphics[width=0.5\textwidth,clip=]{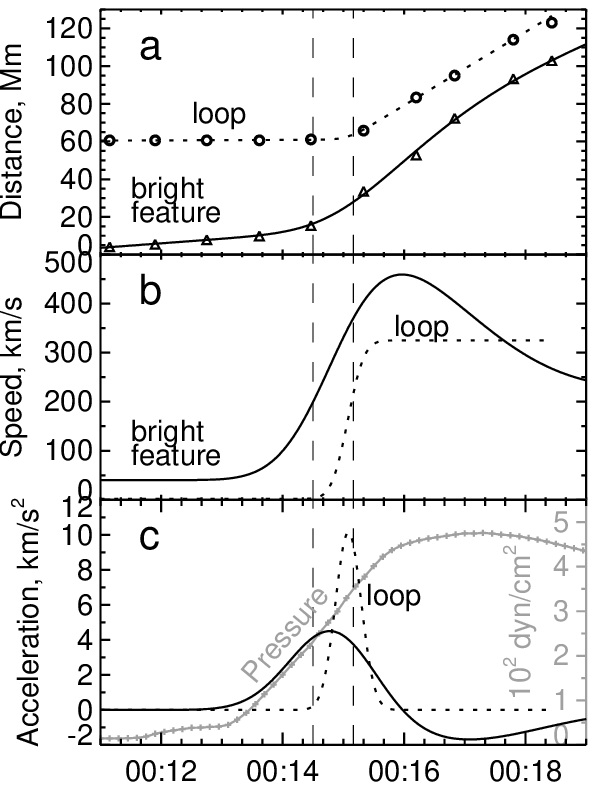}
               \includegraphics[width=0.5\textwidth,clip=]{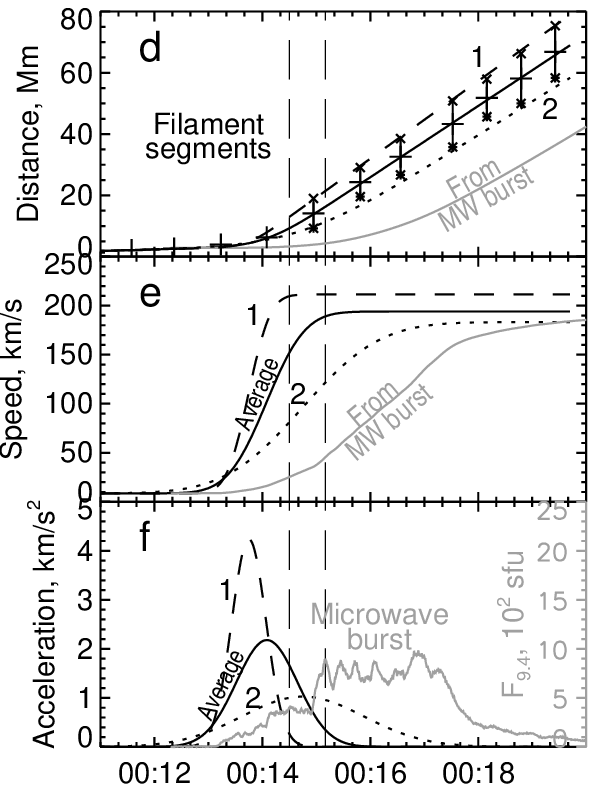}
              }
\caption{Kinematical plots of the eruptions in the 13 July 2004
event: the loop and bright feature (left) and the filament
segments 1 and 2 (right; the solid curves correspond to averages
between 1 and 2). Symbols mark the measured plane-of-sky
distances, and curves represent their fit. The gray curve in panel
(c) displays the plasma pressure computed from GOES SXR fluxes.
The gray curves in panels (d\,--\,f) show kinematical plots
calculated by assuming the correspondence of the acceleration plot
to the microwave one. The vertical dashed lines delimit the start
time of the wave estimated by Grechnev \textit{et al.} (2008).
        }
   \label{F-20040713_kinematics}
   \end{figure}

The results of plane-of-sky measurements are shown in
Figure~\ref{F-20040713_kinematics} for the loop and bright feature
(left) and for the filament segments (right). The dashed curve in
Figure~\ref{F-20040713_kinematics}d was calculated from the dashed
acceleration profile in Figure~\ref{F-20040713_kinematics}f to
match an upper envelope of the measured positions of segment 1.
The dashed curve in Figure~\ref{F-20040713_kinematics}e is a
corresponding velocity profile. The dotted curves in
Figure~\ref{F-20040713_kinematics}d\,--\,f correspond to a lower
envelope of the positions of segment 2. The black solid curves
correspond to an average height\,--\,time plot. The horizontal
bars show the exposure durations. We consider the maximum
acceleration of the middle part of the filament to reach about
2~km~s$^{-2}$ ($-60\%, +120\%$), and its probable speed and
positions to be close to the solid black curves, being within the
dotted and dashed boundaries. The bright feature and eruptive loop
are visible much better, their leading edges are well defined, and
therefore uncertainties of their positions are significantly less.
The estimated height\,--\,time profiles match the measurements.
The main uncertainty here is due to temporal undersampling, which
is only crucial for the estimate of acceleration of the loop.

When the measurements began, the bright feature already rose that
probably corresponded to the initiation phase, which started at
about 00:07 according to soft X-ray (SXR) GOES data. A strong
acceleration started at about 00:13, reached $\approx
4$~km~s$^{-2}$, and then changed to deceleration. The loop was
static by 00:14:30; after 00:15:10 its speed sharply changed to
$V_{\mathrm{loop}\,\max} \approx 320$~km~s$^{-1}$ and did not
increase afterwards. The last measured point hints at a possible
later deceleration. The transition from the initial zero speed to
a final one occurred between two samples. Hence, the maximum
acceleration of the loop could well exceed 10~km~s$^{-2}$. All
components of the corresponding CME decelerated
(\opencite{Grechnev2008}), so that the eruption resembled an
explosion with a strong impulsive acceleration followed by
continuous decreasing deceleration.

Figure~\ref{F-20040713_kinematics}c also shows the plasma pressure
computed from SXR GOES fluxes and a source size of 15~Mm found
from RHESSI \cite{Lin2002} images. The pressure gradually rose
while the bright feature suddenly started to decelerate. Thus, the
flare pressure was unlikely a driver of either the eruption or the
wave, whose estimated start time is delimited with vertical dashed
lines.

The filament started to rise nearly simultaneously with the bright
feature. However, both the acceleration and speed of the bright
feature were higher, and it surpassed the filament (see
Figure~\ref{F-20040713-eruptive_system}). The nature of this
feature is difficult to identify. In some images it resembles an
arcade surrounding the filament, but initially it seems to be
located below the filament. This feature might be also one more
filament, which brightened due to heating. An additional
possibility is suggested by a scenario proposed by
\inlinecite{Meshalkina2009}: this feature might be a small-scale
magnetic rope whose eruption destabilized the filament.

The gray curves in Figure~\ref{F-20040713_kinematics}d\,--\,f show
the kinematical plots calculated under the assumption that the
acceleration plot corresponded to the 9.4~GHz light curve
(Nobeyama Radio Polarimeters, \opencite{Torii1979};
\opencite{Nakajima1985};
\url{ftp://solar.nro.nao.ac.jp/pub/norp/xdr/}). Microwave bursts
are known to be close in shape to hard X-ray (HXR) ones, while
accelerations of eruptions have been found to be close to HXR
bursts (\textit{e.g.}, \opencite{Temmer2008}). The plots
calculated from the microwave burst lag behind the actual plots of
the filament by about two minutes indicating that, most likely,
the flare was caused by the eruption. This fact suggests that the
eruptive filament accelerated almost independently of the flare
reconnection rate and HXR emission, at least, in this event.

There are two options regarding a relation between the bright
feature and the loop. One possibility is that $\approx 1.5-2$
minutes after the start of the acceleration of the bright feature,
the loop suddenly and independently underwent much higher
impulsive acceleration. Alternatively, the loop was expelled by a
shock front that appears to be more probable. For the latter case,
the strength of the shock can be estimated. The Mach number is $M
= V_{\mathrm{sh}}/V_{\mathrm{fast}}$, where $V_{\mathrm{sh}}$ is a
shock speed, and $V_{\mathrm{fast}}$ is a fast-mode speed. The
shock speed $V_{\mathrm{sh}}$ at the onset of the loop motion can
be roughly estimated from a PL fit \cite{Grechnev2008} to be about
1000~km~s$^{-1}$, but it is rather uncertain because of
insufficient temporal coverage by TRACE images and their long
exposure times. The fast-mode speed can be estimated from an
expression $V_{\mathrm{sh}} \approx V_{\mathrm{fast}} + \kappa
U_{\mathrm{sh}}/2$, where $\kappa$ is a coefficient determining
the steepening rate of the wave front. This coefficient, $1/2 \leq
\kappa \leq 3/2$, depends on plasma beta and the propagation
direction. We take the speed of the loop as the gas speed behind
the shock front, $U_{\mathrm{sh}} = V_{\mathrm{loop}\,\max}
\approx 320$~km~s$^{-1}$ (see
Figure~\ref{F-20040713_kinematics}b). With these quantities and
$\kappa \approx 3/2$ for the wave propagation perpendicular to the
magnetic field in low-beta plasma, the Mach number is $ M \approx
1.3$. One might suppose that the steepening time was about the
interval between the peaks of the solid and dotted curves in
Figure~\ref{F-20040713_kinematics}c corresponding to the
acceleration profiles of the bright feature and the loop,
\textit{i.e.}, about 0.5 min.

We extrapolated coronal magnetic fields from a SOHO/MDI
magnetogram using a package of \inlinecite{RudenkoGrechnev1999}
based on a potential approximation \cite{Rudenko2001}. The result
showed that the eruptive loops were strongly tilted to the
photosphere, and their tops were at a height of about 30 Mm. With
a height of the pre-eruptive filament of $> 10$ Mm, the shock
front hit the loop almost horizontally, and the estimated Mach
number is related to the horizontal direction, while in the
vertical direction the shock was probably stronger.

Thus, the loop itself was most likely passive, and its motion was
driven by the shock wave. All observed products of the eruption
monotonically decelerated starting from TRACE observations and up
to LASCO/C3 ones. The loop therefore is unlikely to have excited
the second shock wave, as PHS hypothesized.

\subsubsection{Moreton/EUV Wave, Type II burst, and CME}

\begin{figure} 
  \centerline{\includegraphics[width=0.6\textwidth]
   {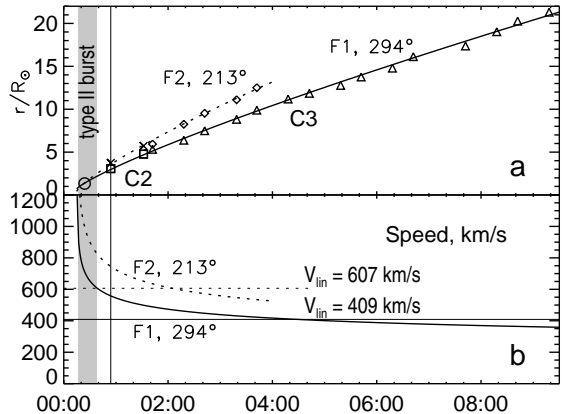}
  }
  \caption{a)~Measurements from the CME Catalog for the F1 and F2
CME components fitted with the shock-PL model, $\delta \approx
2.6$. The circle denotes the position of the off-limb EUV wave at
$294^{\circ}$. b)~Speed\,--\,time plots corresponding to each fit.
Linear-fit speeds are also specified. Shading denotes the interval
in which the type II burst was observed.}
  \label{F-20040713_cme_fit}
  \end{figure}

\inlinecite{Grechnev2008} could not find out from kinematics if
the leading edge of a coronal transient observed by LASCO (see
Figure~\ref{F-20040713-eit_lasco}b) was a mass ejection or a trace
of a wave. Comparison with a non-subtracted image of the corona
before the CME in Figure~\ref{F-20040713-eit_lasco}c suggests its
spiky leading fringe to be coronal rays deflected by a wave
(pronounced for feature F1 at $\mathrm{PA} = 294^{\circ}$).
\inlinecite{Vourlidas2003} and \inlinecite{Magdalenic2008}
interpreted such phenomena as signatures of shocks. Assuming that
the CME leading edge was due to the wave, we outlined the
measurements from the CME Catalog with a shock-PL fit. The
calculated curves with $\delta \approx 2.6$ corresponding to the
Saito model agree with the measurements
(Figure~\ref{F-20040713_cme_fit}). The decreasing speeds computed
from these height\,--\,time plots at distances corresponding to
the LASCO/C2 and C3 fields of view agree with the linear-fit
speeds estimated in the CME Catalog, while within the interval
when the type II burst was observed the speeds are higher by a
factor of $2-2.5$.

\begin{figure} 
  \centerline{\includegraphics[width=0.7\textwidth]
   {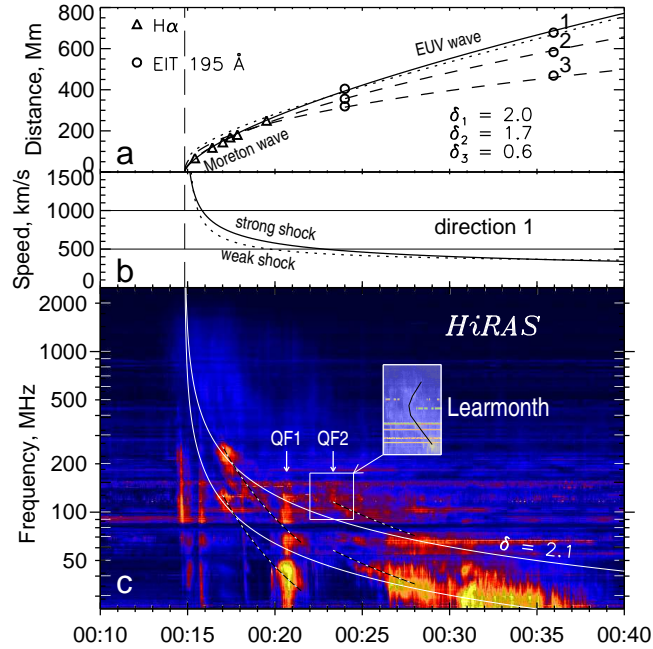}
  }
  \caption{The 13 July 2004 event. Height\,--\,time (a) and
velocity\,--\,time (b) plots of the Moreton and EUV waves, and (c)
the fit of the type II burst ($\delta = 2.1$). Qf1 and QF2 are
features open to question. A reverse drift of Qf2 is detectable in
the inset which shows a portion of the dynamic spectrum recorded
in Learmonth with a higher spectral resolution.}
  \label{F-20040713_type_II}
  \end{figure}

Figure~\ref{F-20040713_type_II}c shows a type II burst recorded by
the HiRAS radio spectrograph and our shock-PL fit of both the
fundamental and second-harmonic emission (solid lines). We use the
density falloff index $\delta = 2.1$, which is close to the
Newkirk model expected for a streamer. Features QF1 and QF2 are
open to question; their presence hinted PHS at two different shock
waves. The dotted lines approximately reproduce the outline of PHS
following the logic suggested by their Figs. 4 and 6. They
correspond to fixed velocities of the type II exciters. However, a
flare blast wave proposed by the authors is expected to
decelerate: just our outline corresponds to a freely propagating
blast wave.

The solid shock-PL fit outlines the whole slowly drifting
structure from the decimetric range up to the lowest frequency. A
question remains about features QF1 and QF2. The former feature
with an uncertain harmonic structure does not seem to favor the
dotted outline relative to the solid one. The weaker QF2 feature,
which PHS considered as the onset of the second type II burst,
indeed seems to have a harmonic structure. Its shape in the
higher-resolution spectrum recorded at the Learmonth station (US
Air Force RSTN) resembling ``('' outlined with a black arc
suggests an encounter of a shock wave with a dense structure (see
Section~\ref{S-shock_type_II}). These facts support association of
the type II burst with a single decelerating shock wave. The
drifting continuum, which PHS found to start at 00:13 (confirmed
by the acceleration profile in
Figure~\ref{F-20040713_kinematics}f), might be due to emission
from outside of a pre-shock region expanding towards a decreasing
density or, alternatively, from inside of the expanding region
with a progressively depleting density, as PHS proposed.
Compression of the environment in the pre-shock interval from
00:13 to about 00:14:50 might have produced an excessive plasma
density, which we describe with a radial power-law falloff.

For comparison we show in Figure~\ref{F-20040713_type_II}a the
distance\,--\,time measurements of the Moreton wave (triangles)
and the EUV wave (open circles) from \inlinecite{Grechnev2008}
along with shock-PL plots for the three directions 1, 2, and 3
denoted in Figure~\ref{F-20040713-eit_lasco}a, and the dotted plot
of a spherical weak shock propagating in uniform plasma calculated
by using expressions of \inlinecite{Uralova1994}.
Figure~\ref{F-20040713_type_II}b presents the speeds for the
strong (solid) and weak (dotted) shock approximations along
direction 1. Both approximations are close to each other far from
the eruption center ($r \gsim R_1 \approx 200$ Mm) being somewhat
different at shorter distances. The shock wave propagating along
the surface probably became weak at $r \gsim R_1$, when it left
the active region and entered quiet Sun's areas where the coronal
density and the fast-mode speed were nearly constant,
\textit{i.e.}, $\delta \to 0$. Closer to the eruption site ($r <
R_1$), the plasma density presumably had a power-law falloff, and
the shock was not weak. These conditions seem to favor the
self-similar shock approximation, in which \textit{i})~the shock
propagation speed is proportional to its intensity and does not
depend on the fast-mode speed, and \textit{ii})~the wavelength is
equal to the distance passed by the wave, \textit{i.e.}, its
duration increases. The density falloff of $\delta <3$ within an
active region and at its periphery corresponds to deceleration and
damping of such a shock. The limit of a strong shock is a
convenient idealization to describe the formation stage of a
single shock wave, which propagates far from its source. A real
forming shock wave does not seem to be so strong that the decrease
of the fast-mode speed could be neglected when the shock leaves
the active region. This issue is beyond our scope. We only note
that formation of the shock discontinuity in a disturbance
produced by an impulsive piston presumably completes (and its
intensity reaches maximum) in a region where the falloff of the
fast-mode speed is steepest.

Note that deceleration of the EUV wave sweeping over the quiet
solar area was stronger towards the equator, as comparison of the
three fronts in Figure~\ref{F-20040713-eit_lasco}a for the
1\,--\,3 directions shows. This is expected for a strong shock,
whose deceleration is determined by the density distribution,
which is maximum at the equator (see the Saito model). This is
also expected for a weak shock, whose propagation is governed by
the Alfv{\' e}n velocity decreasing towards the equator due to
both the density distribution and the dipole magnetic field of the
Sun.

Our analysis of Event 1 has revealed a probable excitation of a
single wave by an impulsively accelerated eruption and steepening
into a shock within one minute. Then the wave freely propagated
like a decelerating blast wave and probably formed the leading
edge of the CME. Our results are consistent with the conclusions
of PHS about the role of a rapidly expanding eruption, formation
of the shock wave at a very low altitude, and their estimates of
the shock speed.

\subsection{Event 2: 1 June 2002}
 \label{S-event2}

\begin{figure} 
  \centerline{\includegraphics[width=\textwidth]
   {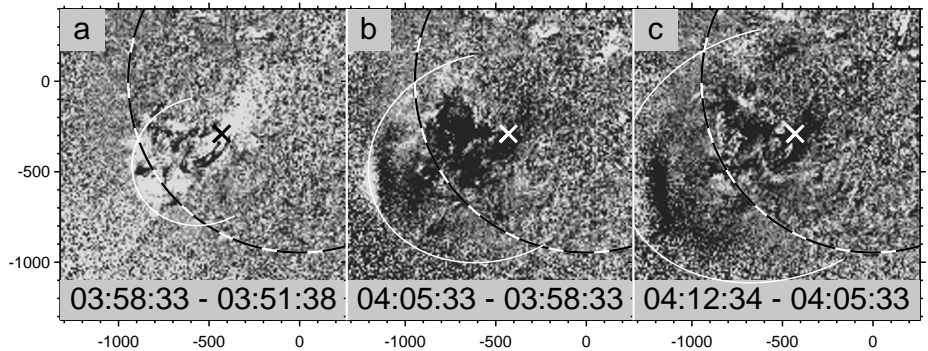}
  }
  \caption{An EUV wave (outlined with ovals) on 1 June 2002 in
EIT 195~\AA\ running-difference images. The dashed circles denote
the solar limb. The slanted cross marks the flare site.}
  \label{F-2002-06-01_euv_wave}
  \end{figure}

\inlinecite{Meshalkina2009} revealed a possible coronal shock wave
presumably excited by a collision of an eruptive magnetic rope
with a magnetic obstacle in the 1 June 2002 event. An M1.5 flare
(S19~E29) started at 03:50 and lasted only 11 min. SOHO/EIT
carried out the `High cadence 195' program, and LASCO did not
observe at that time. Figure~\ref{F-2002-06-01_euv_wave} shows an
off-limb EUV wave in this event.

\begin{figure} 
  \centerline{\includegraphics[width=\textwidth]
   {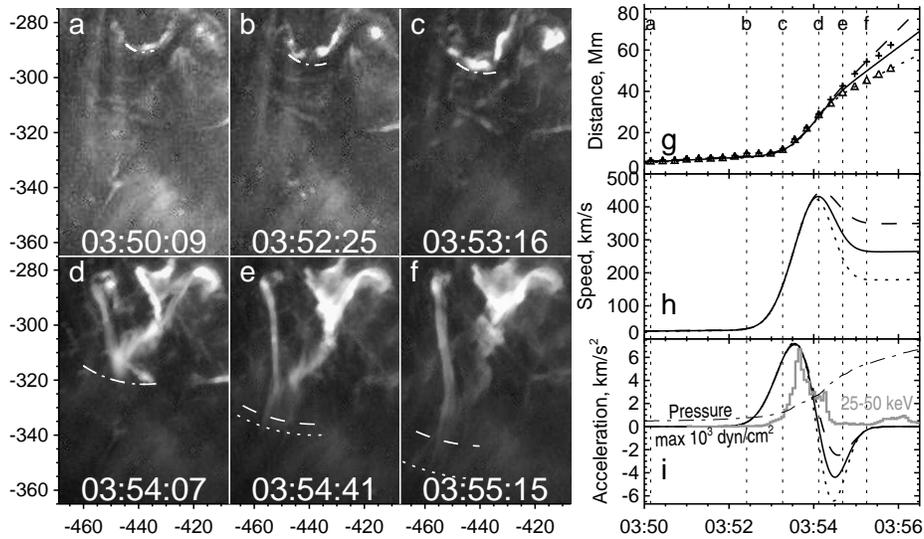}
  }
  \caption{The eruption in TRACE 195~\AA\ images (a\,--\,f) and
its kinematics (g\,--\,i). The dotted and dashed lines correspond
to extreme outlines of the eruption. Panel (i) also shows a
25\,--\,50 keV RHESSI time profile (gray) and the plasma pressure
calculated from GOES data (dash-dotted). The vertical dotted lines
mark the observation times of images a\,--\,f.}
  \label{F-20020601_trace}
  \end{figure}

Figure~\ref{F-20020601_trace}a\,--\,f presents the eruption with
extreme outlines (dotted and dashed) of its leading edge found
from two sets of images processed in different ways.
Figure~\ref{F-20020601_trace}g\,--\,i presents plane-of-sky
measurements of kinematics using the same technique as for
Event~1. The eruption accelerated up to $\approx 7$ km~s$^{-2}$
and then decelerated. The deceleration might be overestimated,
because the eruption started to disintegrate and become
transparent. Similarly to Event 1, acceleration occurred during
the rise of the HXR burst recorded with RHESSI, while the plasma
pressure gradually increased all the time.

\begin{figure} 
  \centerline{\includegraphics[width=0.7\textwidth]
   {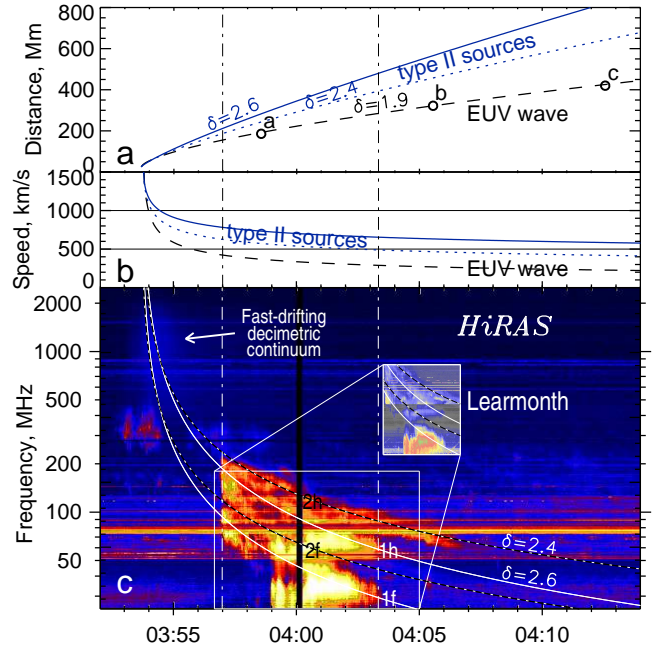}
  }
  \caption{The 1 June 2002 event. Height\,--\,time (a) and
velocity\,--\,time (b) plane-of-sky plots of the EUV wave (black)
and the fit of the type II burst source (blue). Open circles in
panel (a) represent the distances of the foremost wave fronts from
the eruption center, and labels `a'\,--\,`c' correspond to the
images in Figure~\ref{F-2002-06-01_euv_wave}. Panel (c) shows the
dynamic spectrum with a shock-PL fit. The vertical broken lines
delimit the interval of the type II burst.}
  \label{F-20020601_type_II}
  \end{figure}

Figure~\ref{F-20020601_type_II} shows kinematics of the EUV wave
(a,b) and the type II burst (c) similarly to
Figure~\ref{F-20040713_type_II}. To reveal the harmonic structure
of the burst, we use again the record made in Learmonth (the
inset). The burst consisted of two pairs of emission bands with
frequency ratios in pairs of 2.0 and $\approx 1.5$ between the
pairs. The two pairs of bands resemble band-splitting usually
interpreted as the plasma radiation from the regions upstream and
downstream of a shock. However, the relative split is atypically
large for the metric range \cite{Vrsnak2001}. Alternatively, this
situation suggests propagation of the shock front along two
streamers located close to each other. We outline the burst
structure with two harmonic pairs of power-laws 1 and 2 with `f'
indicating the fundamental emission and `h' the second harmonic.
The difference in $\delta$ (2.4 and 2.6) might be due to
differences of density falloffs in the streamers as well as
different angles between the shock front and the axes of the
streamers. The estimated wave onset time is 03:53:40, close to the
acceleration peak time (Figure~\ref{F-20020601_trace}i).

Figure~\ref{F-20020601_type_II}a shows height\,--\,time plots
corresponding to the drift of the type II burst along with the EUV
wave expansion. The type II emission was observed when its source
was presumably located at heights from 190 Mm (heliocentric
distance of $1.27 R_{\odot}$) to 500\,--\,600 Mm [$(1.7-1.9)
R_{\odot}$], i.e., lower than usually assumed. Moreover, the
outline of the shock front matches the fast-drifting decimetric
continuum suggesting its relation to the shock wave, which
presumably appeared still lower. The type II burst probably
started when the shock front reached the streamer and ceased due
to deceleration and damping of the wave.

\subsection{Event 3: 19 May 2007}
 \label{S-20070519}

This event associated with a B9.5 flare at 12:48\,--\,13:19
(N07~W06) and a fast CME has been well studied due to efforts of
several researches mostly from observations made with EUV Imager
(EUVI) of SECCHI complex \cite{Howard2008} on STEREO
\cite{Kaiser2008}. Nevertheless, some questions remain.

\inlinecite{Long2008} measured kinematics of the EUV wave, found
distinct deceleration, and stated that the low velocities of ``EIT
waves'' could be due to their temporal undersampling. The
observations were found to be consistent with an impulsively
generated fast-mode magnetosonic wave or propagating MHD shock.
However, they revealed an initial acceleration of the disturbance
from a nearly zero speed. This mismatches an expected behavior of
an MHD wave.

\inlinecite{VeronigTemmerVrsnak2008} also found deceleration of
this disturbance indicative of a freely propagating MHD wave and
revealed a wave reflection at a coronal hole. They assumed that
the wave was initiated by the CME, because the associated flare
was very weak and occurred too late to account for the wave
initiation. They also revealed two eruptions following each other.

\inlinecite{Gopalswamy2009} measured propagation of reflected wave
fronts and considered the reflections as an argument in favor of a
wave nature of EUV transients (see also \opencite{Schmidt2010}).
However, \inlinecite{Attrill2010} proposed that the reflections
resulted from a misinterpretation of the running difference data
and suggested instead that two EUV wave fronts developed during
the event. Indeed, running differences reliably show only the
outer boundaries of expanding disturbances, while the inner
picture reflects all changes occurring between two images
subjected to subtraction \cite{ChertokGrechnev2005}.

  \begin{figure} 
  \centerline{\includegraphics[width=\textwidth]
   {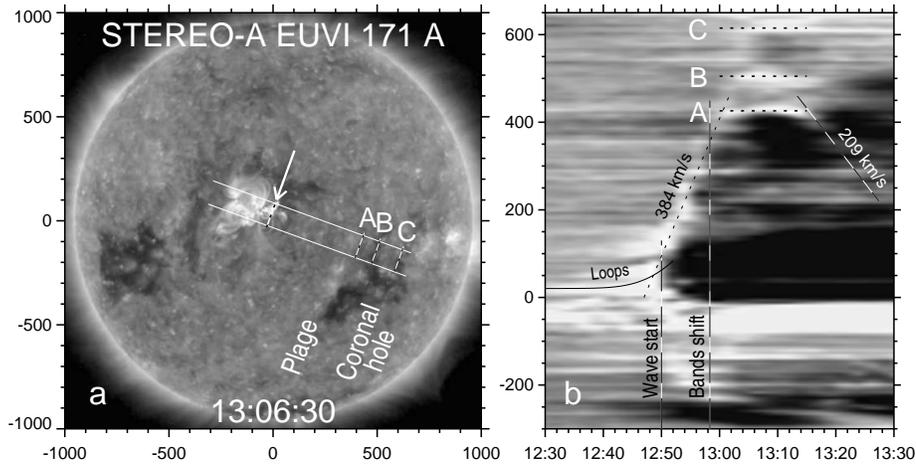}
  }
  \caption{EUV wave observed with STEREO-A/EUVI at 171~\AA\ on 19 May 2007.
A set of spatial profiles (b) computed from ratio images within
the strip contoured in panel (a). The arrow points at the origin
of measurements. The A, B, and C broken bars in both panels mark
the reflection positions. The slanted broken lines in panel (b)
outline the steepest slopes. }
  \label{F-20070519_euvi}
  \end{figure}

To see whether or not the reflections actually occurred, we use a
movie composed of non-subtracted 171~\AA\ images of STEREO-A/EUVI
(euvi\_ahead\_171.gif in the electronic version of the paper). A
backward motion suggesting a reflection is visible after 13:11
northeast from a plage region denoted in
Figure~\ref{F-20070519_euvi}a. We repeated the measurements of
\inlinecite{Gopalswamy2009}, but without any subtraction. We only
consider the first probable reflection in the direction exactly
backwards, because it is difficult to reveal wave fronts in
non-subtracted images of this complex event.
Figure~\ref{F-20070519_euvi} corresponds to Figs. 1 and 2 from
their paper. To enhance the sensitivity, we use in panel (b)
spatial profiles computed as the sums over the width of each slice
extracted from an image as panel (a) shows. Each image was
normalized to a pre-event image at 12:16:30 (fixed-base ratios).

Figure~\ref{F-20070519_euvi}b shows that the plage region
inflected after arrival of the wave front and then returned back.
A backwards motion from plage A is faintly visible after 13:14.
Region B exhibited a weaker bend. It is not clear from which
region of A and B was the wave reflected. The latter seems to be
preferable, because continuations of both the direct and reflected
slanted traces intersect farther from region A, while a prolonged
standing of the wave at region A is doubtful. The slanted broken
lines represent the speeds found by \inlinecite{Gopalswamy2009}
and agree with the slopes in Figure~\ref{F-20070519_euvi}b. Thus,
the results of the authors were correct, at least, for the first
reflection. The wave reflected backwards was considerably slower
than the incident wave. This fact supports the shock-wave nature
of the disturbance. Indeed, if an incident shock wave propagating
with a velocity $V_{\mathrm{inc\ sh}}$ encounters a
semitransparent `wall' like a coronal hole, then the shock
reflected backwards is slower: $V_{\mathrm{back\ sh}} \approx
V_{\mathrm{inc\ sh}} - V_{\mathrm{gas}}$, where $V_{\mathrm{gas}}$
is a velocity of the gas trailing the incident shock front.
$V_{\mathrm{gas}}$ can be up to the sound speed.

\begin{figure} 
  \centerline{\includegraphics[width=\textwidth]
   {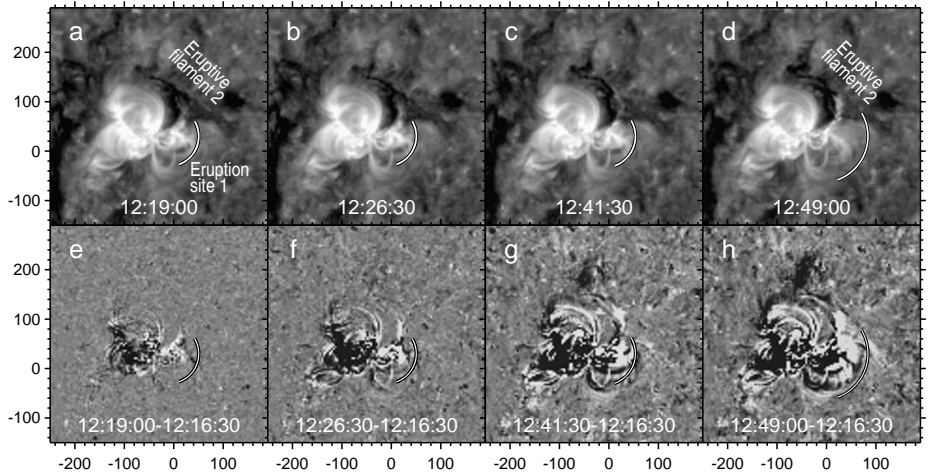}
  }
  \caption{The first eruption on 19 May 2007 observed with STEREO-A/EUVI at 171~\AA.
Top: non-subtracted images, bottom: fixed-base differences. The
arcs outline the foremost edge of the eruptive loop system
according to the fit shown with the solid line in
Figure~\ref{F-20070519_euvi}b.
  }
  \label{F-20070519_eruption}
  \end{figure}

What does the accelerating part prior to 12:50 display?
Figure~\ref{F-20070519_eruption} shows EUVI 171~\AA\ images with
subtraction and without it. The outermost boundary of the
expanding bright feature coincides with the edge of coronal loops
visible in the earlier non-subtracted images. Then eruptive loops
rapidly loose brightness due to expansion and become invisible in
non-subtracted images. We conclude that the accelerating part
measured by \inlinecite{Long2008} was related to the expanding
loops, while the decelerating part was related to the wave. It is
difficult to distinguish an appearing wave, which brightens, from
a piston (loops), which becomes transparent. For this reason the
acceleration of the loops is uncertain within
$120-270$~m~s$^{-2}$; the outline in Figures~\ref{F-20070519_euvi}
and \ref{F-20070519_eruption} corresponds to $260$~m~s$^{-2}$.

\begin{figure} 
  \centerline{\includegraphics[width=\textwidth]
   {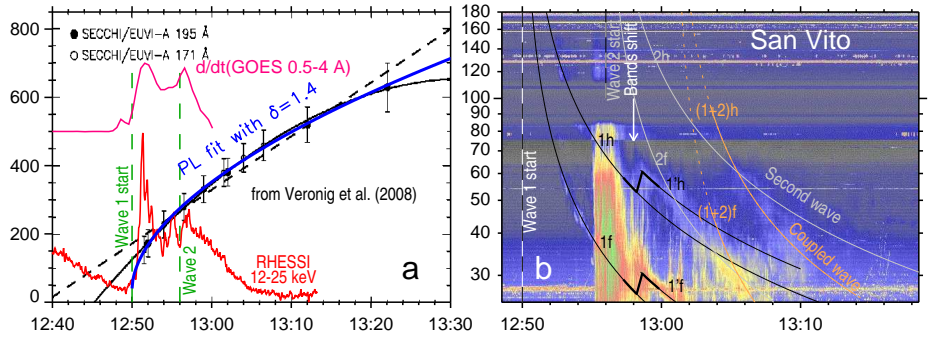}
  }
  \caption{The 19 May 2007 event. a)~Propagation of the EUV wave
measured by Veronig, Temmer, and Vr{\v s}nak (2008, black) along
with an HXR time profile (red), a derivative of the SXR flux
(pink), and a shock-PL fit (blue). b) Type II burst with
shock-PL-fitted four pairs of bands. Paired bands are shown with
the same colors. See details in the text.}
  \label{F-20070519_wave_spectrum}
  \end{figure}

Figure~\ref{F-20070519_wave_spectrum}a shows the measurements of
the wave presented by \inlinecite{VeronigTemmerVrsnak2008} and
their shock-PL fit (thick blue curve). The wave start (12:50)
corresponds to the early rise phase of the HXR burst (red, also
from their paper). The power-law fit corresponds to the
measurements of the authors better than their quadratic fit and
the linear one. The eruption in this event also accelerated before
the appearance of manifestations of flare reconnection.

The suggestion of \inlinecite{Attrill2010} about the second wave
appears to be correct. The second eruption was probably triggered
by the first one. Filament 2 activated at about 12:47 and erupted
at 12:55\,--\,12:57 according to TRACE 173~\AA\ images. The HXR
time profile was complex, but two distinct episodes are detectable
in the derivative of the SXR flux recorded with GOES (pink in
Figure~\ref{F-20070519_wave_spectrum}a). The onset times of the
two waves were about 12:50 and 12:56.

\begin{figure} 
  \centerline{\includegraphics[width=\textwidth]
   {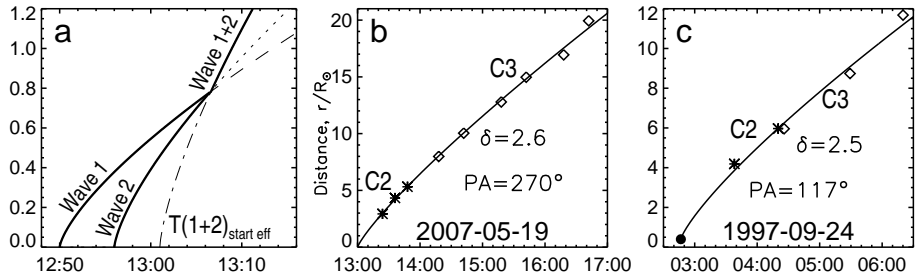}
  }
  \caption{Coalescence of two shock waves (a) and
height\,--\,time plots of CMEs observed on 19 May 2007 (b) and 24
September 1997 (c). Symbols present data from the CME Catalog,
lines show their shock-PL fit. The filled circle in panel (c)
denotes the origin of the wave.}
  \label{F-cartoon2_cme}
  \end{figure}

A shock wave trailing a preceding one must reach the leading front
due to properties of shock waves; the two shock waves coalesce to
produce a single shock front (and a weak backwards disturbance,
which we are not interested in). Its speed is less than the sum of
the initial fronts' speeds; however, the resulting shock is
stronger and faster than either of the initial ones. So the slope
of its distance\,--\,time plot is steeper than the initial waves
had, and its virtual onset time is later than for either of the
initial waves, as Figure~\ref{F-cartoon2_cme}a outlines.

Let us try to understand a complex dynamic spectrum in
Figure~\ref{F-20070519_wave_spectrum}b recorded in San Vito (USAF
RSTN). A harmonic pair of weak bands 1f,\,1h is sometimes
detectable after 12:52. A strong type III burst at
12:55\,--\,12:58 probably associated with the second eruption
partially blocked the weaker bands 1f,\,1h. Two stronger type II
bands 1'f,\,1'h appeared at 12:58. Their appearance looks like an
inverse-N-like shift of the initial bands to higher frequencies
(thicker outline in Figure~\ref{F-20070519_wave_spectrum}b). The
initial bands 1f,\,1h are still detectable sometimes. The outline
of both these pairs of bands has the same start time of the first
wave, 12:50. The appearance of the second pair might be due to the
entrance of a part of the shock front into a dense region located
rather high above plage A (see Figure~\ref{F-20070519_euvi}a,b).
The surface EUV wave reached plage A slightly later, as expected
for a convex front tilted towards the solar surface (see
Section~\ref{S-shock_type_II}).

The second shock front was probably manifest in weak bands 2f,\,2h
(pale, the start time is 12:56). The bands denoted (1+2)f and
(1+2)h (orange outline) probably reveal the resulting shock with a
virtual start time of 13:01. All of these bands overlap with
others increasing the total emission at the intersections. We
remind that various bands were most likely emitted from spatially
different sites.

We have not revealed manifestations of the two merging shock waves
in images or in the kinematical plot. This result could be
expected due to different propagation conditions for a shock wave
upwards and along the solar surface. The near-surface portion of a
shock front decelerates stronger and experiences significant
damping, thus becoming weak and approaching a linear disturbance.
If the trailing shock wave succeeded to catch up or intersect with
the leading one, the effect is expected to approach an interaction
of linear disturbances, when two waves pass through each other
experiencing interference, while the scheme in
Figure~\ref{F-cartoon2_cme}a shows an essentially nonlinear
effect.

The CME was fast (958~km~s$^{-1}$) and decelerated. We assume that
its leading edge was a trace of a wave and apply a shock-PL fit to
the measurements in the CME Catalog at $\mathrm{PA} = 270^{\circ}$
(Figure~\ref{F-cartoon2_cme}b). The onset corresponds to the
virtual start time of the coupled shock wave, and the density
falloff corresponds to the Saito model. A trailing poorly observed
CME at $\mathrm{PA}  = 310^{\circ}$ and probably related to the
same event was considerably slower (294~km~s$^{-1}$). Its
deceleration might be due to the influence of the wave running
ahead and problems of measurements.

Our analysis does not pretend to be perfect, but it shows that
even such a complex dynamic spectrum can be reconciled with EUV
observations and CME expansion under assumption of the shock-wave
nature of related disturbances. Oversimplified considerations of
such a complex event can be misleading. In particular,
\inlinecite{YangChen2010} concluded that the EUV wave in this
event ran slower in regions of stronger magnetic field that seemed
to be a challenge for the wave hypothesis. However, 1)~the authors
considered the radial component of the magnetic field only,
whereas the Alfv{\'e}n speed depends on its magnitude. 2)~By
taking the range of the magnetic field strengths $ \leq 0.6$~G,
for which \citeauthor{YangChen2010} obtained anticorrelation with
the EUV wave speed, and a density of $ \gsim 2 \times
10^{8}$~cm$^{-3}$ from the Saito model, one obtains $\beta =
(2/\gamma) C_\mathrm{s}^2/V_\mathrm{A}^2 \gsim 4$, \textit{i.e.},
the wave must be almost insensitive to the magnetic field. 3)~The
fronts in their Fig.~4 stretch west-southwest and become sharply
pointed at 12:59, whereas \inlinecite{Long2008},
\inlinecite{VeronigTemmerVrsnak2008}, and \inlinecite{Attrill2010}
showed the front to be blunt in this direction at that time.
4)~Usage of the Huygens plotting to find trajectories of the wave
front resulted in a strange picture of intermittently condensed
and rarefied ray trajectories in their Fig.~4. Thus, the results
of \inlinecite{YangChen2010} do not offer problems for the
shock-wave hypothesis.

\subsection{Event 4: 24 September 1997}

\begin{figure} 
  \centerline{\includegraphics[width=\textwidth]
   {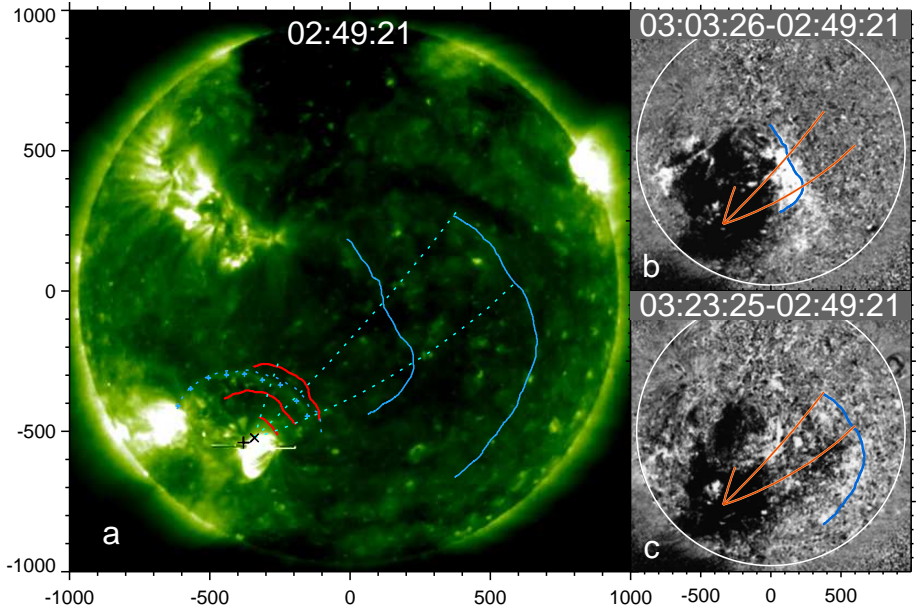}
  }
  \caption{EUV wave (blue) and Moreton wave in the 24 September
1997 event. a)~Non-subtracted EIT 195~\AA\ image with the first
front (blue crosses). Blue dotted line outlines its foremost edge.
Red fronts outline the Moreton wave. White broken lines trace the
measurement great circles. Black crosses denote the origins of
measurements. b,c)~Difference ratio images with blue outlines of
EUV wave fronts and red measurement great circles. The outlines
correspond to Warmuth \textit{et al.}, 2004a (courtesy A.
Warmuth). }
  \label{F-19970924_eit}
  \end{figure}

This event was associated with a short M5.9 flare
(02:43\,--\,02:52, S31~E19). An H$\alpha$ Moreton wave and EUV
wave in this event were first analyzed by
\inlinecite{Thompson2000}. \citeauthor{Warmuth2004a}
(\citeyear{Warmuth2004a,Warmuth2004b}) found kinematical closeness
of both wave fronts to each other and their deceleration. The
first EUV wave front (Figure~\ref{F-19970924_eit}a) was sharp and
bright suggesting that the main EUV-emitting layer was low. A
difference ratio image (Figure~\ref{F-19970924_eit}b) reveals weak
wave manifestations south, southwest, and slightly west from the
outline of \inlinecite{Warmuth2004a}. The third front in
Figure~\ref{F-19970924_eit}c is close to their outline.
Deceleration of the EUV wave was therefore even slightly stronger
than the authors estimated.

\begin{figure} 
  \centerline{\includegraphics[width=\textwidth]
   {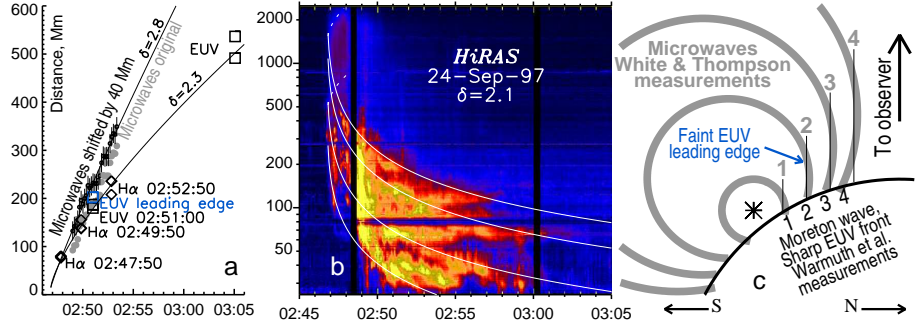}
  }
  \caption{a)~Propagation of the EUV wave (squares) and Moreton
wave (diamonds) measured by Warmuth \textit{et al.} (2004a).
Filled gray circles show the of White and Thompson (2005) data,
black open circles with error bars show them shifted by 40 Mm.
Both data sets are outlined with shock-PL fit. b)~The HiRAS
dynamic spectrum outlined with shock-PL fit and a negatively
drifting continuum (dotted). c)~Presumable relation between the
layers emitting microwaves and EUV/H$\alpha$. Thin vertical lines
show cross sections of the wave fronts of the largest column
emission measure contributing at 17 GHz. The star is the wave
origin.}
  \label{F-19970924_fit}
  \end{figure}

\inlinecite{WhiteThompson2005} analyzed wave signatures in
microwave images at 17 GHz but did not reveal any deceleration.
They also found: \textit{i})~the speed of the microwave
disturbance was 830~km~s$^{-1}$ against $\approx 500$~km~s$^{-1}$
estimated for the Moreton wave; \textit{ii})~the brightness
temperature at 17 GHz was about five times higher than an estimate
from EIT data, and the discrepancy could be reduced if the kinetic
temperature at 17 GHz would be different (preferentially higher)
from the characteristic temperature of the 195~\AA\ channel. The
authors also concluded that the timing of images should be
corrected by $\approx 100$~s for EIT and by $\sim 180$~s for
H$\alpha$ to reconcile all observations. These facts indicate that
the layers emitting microwaves and EUV were not identical. The
higher speed, lesser deceleration, and higher brightness
temperature (\textit{i.e.}, column emission measure) observed at
17 GHz with respect to EUV hint at a possibly higher location of
the microwave-emitting layer. Figure~\ref{F-19970924_fit} shows
our suggestion in panel (c); panel (a) presents the
distance\,--\,time plots from both papers. The corrected times of
EIT images ($+99$~s) and H$\alpha$ ones ($+170$~s) are specified
at data points.

To reconcile kinematics of the microwave and EUV/H$\alpha$ wave
fronts, we shift the \citeauthor{WhiteThompson2005} data by 40 Mm
and fit both data sets with the same onset time of 02:46:50 but
different power-law exponents. Figure~\ref{F-19970924_fit}c
explains the idea: the lower part of the front propagating in
high-density regions decelerated stronger ($\delta \approx 2.3$
from \citeauthor{Warmuth2004a} measurements), while microwaves
were dominated by long cross sections of the wave front (bars
1\,--\,4) running in lower-density regions, $\delta \approx 2.8$
(\textit{cf.} Paper III). The large-height EUV wave's leading edge
detectable close to the eruption center diminished at large
distances, where the EUV wave was dominated by low structures. The
wave presumably appeared at a considerable height (the star). With
a difference between the origins of measurements (black crosses in
Figure~\ref{F-19970924_eit}a) of $\approx 32$ Mm, our shift of 40
Mm, and spherical wave fronts, we get a height of 117 Mm. It seems
to be overestimated; an estimate of about $\approx 75$ Mm from the
dynamic spectrum appears to be more plausible. Thus, the wave
could be strongly anisotropic starting from its appearance, or the
wave exciter was large, or both.

The complex dynamic spectrum in Figure~\ref{F-19970924_fit}b has a
questionable harmonic structure, which is beyond our scope; we are
only interested in the drift of the burst envelope. We have
outlined presumable fundamental and second-harmonic bands. A
possible higher-frequency band and the highest-frequency envelope
of the burst are formally outlined as the fourth and sixth
harmonic (this does not mean reality of emissions at such high
harmonics). The outline corresponds to the initial height of 75
Mm, the same wave start time of 02:46:50, and $\delta \approx 2.1$
typical of streamers. A negatively drifting continuum (broken
outline) at the initial stage indicates propagation of the shock
front towards the chromosphere.

A poorly observed CME centered at $137^{\circ}$ with a speed of
531~km~s$^{-1}$ was injected into a preceding CME. The CME Catalog
estimates its acceleration to be positive with a remark about
uncertainty. By adding the known origin of the wave (the filled
circle in Figure~\ref{F-cartoon2_cme}c), we get a shock-PL fit of
the measurements from the CME Catalog with an exponent $\delta
\approx 2.5$ close to the Saito model.

Our considerations confirm correctness of both
\inlinecite{Warmuth2004a} and \inlinecite{WhiteThompson2005}
results, reconcile them with each other and with the type II burst
as well as the CME. Deceleration of the front portion detectable
at 17 GHz was much less than its lowest part visible in EUV and
H$\alpha$ had. It was not possible to reveal deceleration from
microwave observations, which allowed detection of the wave within
an interval as short as 4.5 min.

\section{Discussion}
\label{S-Discussion}

TRACE observations of abrupt eruptions in events 1 and 2 have
revealed plane-of-sky accelerations of magnetic rope structures of
$4-7$~km~s$^{-2}$, \textit{i.e.}, (15\,--\,25)-fold gravity
acceleration. Then the eruptions in both events decelerated.
Coronal waves appeared in events 1\,--\,3 approximately at the
peak of acceleration. The onsets of the waves in all four events
corresponded to the rise phases of HXR or microwave bursts (in
event 4 according to \opencite{WhiteThompson2005}). The wave in
event 1 steepened into a shock within one minute and reached a
Mach number of about 1.3 in a horizontal direction, while the
upwards shock could be stronger. In the next Section we consider
which of known shock formation scenarios (see, \textit{e.g.},
\opencite{VrsnakCliver2008}) appear to match the observations.

\subsection{Comparison of Observations with Shock Formation Scenarios}

A scenario of the shock formation by a flare pressure pulse is
based on an idea that the $\beta$ ratio of the plasma pressure to
the magnetic pressure sharply changes from $\beta \ll 1$ to $\beta
> 1$. Such a change of $\beta$ is believed to be possible in a
flare loop and considered as a cause of an omnidirectional
disturbance. Dramatic changes of the volume of a loop or its
abrupt motion are necessary to get a significant intensity of a
wave excited in this way. Objections against this scenario do
exist.

1)~As \inlinecite{Grechnev2006beta} showed, the effect of a high
$\beta$ (even $\beta >1$) in a flare loop is not dramatic, only an
increase of all its sizes of $\sqrt[4]{1+\beta}$. 2)~The idea that
a situation of $\beta \to 1$ can cause instability of a loop is
not confirmed by observations. \inlinecite{IchimotoSakurai1993}
and \inlinecite{Grechnev2006spiders} showed that the $\beta < 1$
condition was not satisfied in long-lived hot coronal loops.
3)~The flare pressures in events 1 and 2 rose gradually, without
any marks of the wave appearance. 4)~RHESSI showed the
SXR-emitting regions in events 1 and 2 to be fixed when the
pressure increased. \inlinecite{VeronigTemmerVrsnak2008} concluded
that the wave ignition in event 3 by the flare was unlikely. Thus,
the theoretical considerations supported by the observational
facts make the ignition of waves by flares doubtful in agreement
with the conclusion of \inlinecite{Cliver2004} (note that the
authors implied the shocks to be driven by outer surfaces of CMEs,
whereas we consider the shocks to appear inside CMEs).

This conclusion is seemingly opposed by the results of
\citeauthor{Magdalenic2008}
(\citeyear{Magdalenic2008,Magdalenic2010}) who presented five
events, in which shocks appeared during the flare impulsive phase,
whereas related CMEs were slow. However, the authors did not
consider a rare type of CME kinematics with very strong initial
acceleration followed by deceleration. An example is our event~1
(Section~\ref{S-event1}, Figure~\ref{F-20040713_kinematics}). The
acceleration stage in this event lasted three minutes. The fastest
feature accelerated up to 470~km~s$^{-1}$ and continuously
decelerated later on. From an overall height\,--\,time plot,
including measurements from LASCO images,
\inlinecite{Grechnev2008} estimated the speed of this feature to
approach 100~km~s$^{-1}$ far from the Sun. Nevertheless, the shock
in this event was most likely excited just by this feature as an
impulsive piston rather than by the flare. Thus, a slower motion
of the main body of a CME well behind a shock front does not
guarantee that the CME or its components were not implicated in
excitation of the shock.

The maximum plane-of-sky speeds of the eruptions in events
1\,--\,3 appear to be well below the Alfv{\' e}n speed expected at
moderate heights ($< 100$ Mm) above active regions (\textit{e.g.},
\opencite{Mann2003}), where the waves appeared. It is possible
that the angles between the velocity vectors and the line of sight
significantly differed from $90^{\circ}$, so that the real
velocities could be higher, but the corresponding corrections are
insufficient to increase the velocities above the Alfv{\' e}n
speed. The time profiles of the velocities estimated for the
eruptions and waves were quite different. Thus, the bow shock
scenario is also unlikely.

\subsection{Impulsive Piston Scenario}
 \label{S-imp_piston_scenario}

In a simplest scheme, a piston moving with a speed $U$ has a
box-like acceleration profile with a value $a$ during the
acceleration phase and zero before and afterwards. An important
condition is that plasma cannot flow around the piston (this
occurs, \textit{e.g.}, in 3D expansion of an arcade). In a flat
geometry and homogeneous medium the plasma flow ahead of the
piston corresponds to a simple wave. The discontinuity appears at
$t_{\mathrm{sh}}^0 = V_{\mathrm{fast}}/(\kappa a)$ at a distance
$r_{\mathrm{sh}}^0 = V_{\mathrm{fast}}t_{\mathrm{sh}}^0$ with $1/2
\leq \kappa \leq 3/2$ that is similar to a solution of an
analogous gas-dynamic problem \cite{LandauLifshits1987}. Then the
speed jump in the discontinuity increases up to the piston's
maximum speed $U_{\max}$. The condition $U > V_{\mathrm{fast}}$
essential for bow shock formation is unnecessary in the impulsive
piston scenario.

Accelerations and their durations before the wave onsets were
(Section~\ref{S-observations}): 4~km~s$^{-2}$ and 90~s in event 1;
7~km~s$^{-2}$ and 70~s in event 2; $0.12-0.27$~km~s$^{-2}$ and
800~s in event 3. The stronger acceleration, the faster a shock
appeared.

The shock waves in the four events were most likely excited by
eruptive structures as impulsive pistons, which one might call the
appearing CMEs. Then the waves rapidly steepened into shocks,
detached the pistons, and freely propagated afterwards like blast
waves. The shock excitation mechanism implies a source height to
be nonzero, but rather low, probably $< 100$~Mm, as suggested by
all dynamic spectra and implied by event~4.
\citeauthor{Magdalenic2008}
(\citeyear{Magdalenic2008,Magdalenic2010}) found the heights of
sources of metric type II bursts to be between 70 and 280~Mm;
however, the apparent heights of limb sources in the metric range
could be noticeably reduced due to refraction (see, \textit{e.g.},
\opencite{Zheleznyakov1970}). The shock character of the waves is
supported by the correspondence of their kinematics to the
expected propagation of shock waves as well as the drift rates of
type II bursts and drifting continua. The shock-wave nature of EUV
waves is supported with probable reflections and coupling of two
shock waves in event 3. Expansion of leading edges of CMEs
produced in three events corresponded to propagation of the lower
skirts of the shock fronts observed as Moreton/EUV waves. Thus,
the wave excitation by an impulsive piston appears to match all
the considered observations, basically corresponding to a picture
proposed by \inlinecite{Uchida1974}.

So far we did not relate a piston with a particular structure or
its surface. Presumable pistons could be either an eruptive
filament (EF) or a CME frontal structure (FS). Both expand as an
entire ensemble in a completely formed CME; only its outer sheath
can be a piston. This sheath is believed to be both the surface of
contact discontinuity and the outer FS surface. Expansion of such
an FS-piston determines propagation of an interplanetary
piston-driven wave and the drag force affecting a CME. The
situation is different during the early CME formation inside an
active region, when EF moves faster than it would be adequate for
a self-similar expansion of the whole CME. The EF acts here as an
impulsive piston and excites inside a future CME a wave, which
freely propagates outwards as a shock wave. In the four events we
revealed just this excitation scenario of waves, which resembled
blast ones. Propagating upwards, such a wave inevitably would pass
through the FS and appear ahead of it.

\subsection{CME Components and Waves}
\subsubsection{Particularities of Expansion}

Expansion of a magnetoplasma CME's constituent is different from
kinematics of a wave traced, \textit{e.g.}, with a leading edge of
a plasma flow driven by a shock. The CME expansion is known to be
about self-similar at moderate distances from the Sun. \textit{The
self-similar approach does not apply to early stages of expansion,
when the structure and shape of a CME have not yet been
established.} When an instability driving an eruption completes
and drag of the solar wind is not yet significant, the
self-similar CME kinematics can be obtained from considerations of
forces affecting a CME \cite{Low1982,Uralov2005}.

Expansion of magnetoplasma structures is governed by magnetic
forces, plas\-ma pressure, and gravity as long as the effect of
the solar wind is small. With the polytropic index $\gamma \approx
4/3$ all the forces integrated over the boundary and volume of a
CME scale with distance $r$ by the same factor of $r^{-2}$. This
leads to an expression for the CME velocity $V_{\mathrm{CME}}^2 =
V_0^2 +(V_{\infty}^2-V_0^2)(1-R_0/r)$, $V_{\mathrm{CME}} = dr/dt$
\cite{Grechnev2008}. Here $R_0$ is the initial size of
self-similar expansion, $V_0$ the initial velocity at $R_0$, and
$V_{\infty}$ the asymptotic velocity in infinity. At large
distances acceleration $\propto r^{-2} \to 0$ and
$V_{\mathrm{CME}} \to V_{\infty}$. The expression for the CME
velocity describes different types of kinematics. The situation
$V_{\infty} \gg V_0 \sim 0$ appears to be typical. Event~1 showed
a different behavior, $V_0
> V_{\infty}$, resembling an explosion with an impulsive
acceleration followed by deceleration. A special type $V_0 \approx
V_{\infty}$ (an impulsive acceleration is required to reach $V_0$)
might correspond to some jet-like ejections, where magnetic
reconnection destroys a structure of an eruption
\cite{Meshalkina2009,Filippov2009,Liu2011}. Event~2 might have
belonged to this type.

Since a FS starts to expand practically from a static equilibrium
in a typical situation $V_{\infty} \gg V_0 \sim 0$, the FS-piston
usually either accelerates or moves with a nearly constant speed
at the self-similar stage. By contrast, shock waves in all
considered events decelerated. Hence, a typical FS-piston is
expected to eventually approach the wave front. What does such a
relation between the speeds of the shock front and piston mean? Do
MHD equations allow a decelerating shock wave to run for a long
time ahead of a non-decelerating FS-piston? We search answers in a
theory developed by \inlinecite{Low1984} in solving a problem of
self-similar expansion of a CME preceded by a strong shock wave.
Though the problem was solved in a limit of very strong shock
propagating in plasma with too steep density falloff $r^{-26/7}$,
the solution correctly relates accelerations of the piston and
piston-driven shock. Assuming a common linear profile of the
plasma velocity in the whole region from the expansion center up
to the shock front, it is possible to relate kinematics of the
shock front and the contact discontinuity, \textit{i.e.}, the
FS-piston without a complete solution of the problem. By fitting
the motion of the piston with a function $r_{\mathrm{pist}} =
bt^m$, we express the sign of the shock acceleration
$a_{\mathrm{sh}}$ vs. $m$: $a_{\mathrm{sh}} \propto (\alpha
m-1)t^{\alpha m-2}$ with $\alpha = (\gamma + 1)/2$. Thus, the
conditions $a_{\mathrm{sh}} < 0$ and $a_{\mathrm{pist}} =
d^2r_{\mathrm{pist}}/dt^2 >0$ are incompatible. Such an FS-piston
and the shock front expand in different ways, which cannot be
coordinated with each other, unlike a bow shock. An impulsively
excited freely propagating shock wave must eventually change to a
piston-driven mode. Presumably this typically occurs at large
distances, probably beyond the LASCO/C3 field of view. The
transformation of a blast shock wave into a piston-driven one
marks switch-on the aerodynamic drag and termination of the
self-similar expansion regime. The drag force becomes significant,
which means establishment of a continuous energy transport from
the FS-piston to the shock wave. By contrast, a blast-like wave
excited by an EF and running ahead of FS, which does not yet act
as a piston, facilitates expansion of a CME into the solar wind.
The shock wave forwards a part of its energy to the FS-piston, and
the drag force is absent.

Most likely, real shock waves are neither purely blast waves nor
purely piston ones. A shock front is sensitive to any events
occurring behind it, \textit{e.g.}, changes of the FS-piston
speed, because the fast-mode speed behind the shock front is
higher than its phase speed. To produce one more shock wave, an
FS-piston has to repeat the maneuver, which produced the first
shock. This is improbable when an instability driving an eruption
has completed and a CME left the Sun.

\subsubsection{Distinguishing between Shock Signatures and CME Components}

\inlinecite{Sheeley2000} and \inlinecite{Vourlidas2003} considered
distortions of coronal streamers as a morphological suggestion of
presumable shocks. Indeed, moving wave-like kinks or deflections
of coronal rays resemble an expected effect of a propagating shock
wave. Some fast ``CMEs'' in difference images might be actually
combinations of coronal rays deflected by shock waves. However,
\inlinecite{Filippov2010} demonstrated that deflection of coronal
rays could be due to expansion of a CME in magnetized corona
without a shock. Irrespective of a particular type of a possible
shock, more reliable morphological suggestion might be a
spike-like leading feature due to deviation of a coronal ray by a
wave. \inlinecite{Magdalenic2008} showed such a situation in their
Fig.~4, where deflected coronal rays were visible well ahead of
trailing CME structures.

We remind that the conic bow-shock shape is not expected, at
least, for wide super-Alfv{\' e}nic CMEs. The shock front must
cling to its foremost edge and closest flanks, while far flanks
and a rear part can be constituted by a freely propagating shock
front, so that the shape of the whole front would resemble an egg.

An attractive way to detect a shock front is to search for
discontinuity in the density distribution shown by coronagraph
images \cite{Vourlidas2003}. However, this way is model dependent.
Besides modeling the coronal density, one has to distinguish
between the shock discontinuity and the contact surface separating
the CME and environment. The 3-dimensional contact surface should
be also modeled. Thus, identifying a shock front in coronagraph
images does not seem to be a simple task. Combinations of
different indications seem to be be useful. A non-structured faint
density enhancement forming the envelope of a transient could be
among them \cite{Vourlidas2003}. Fast decelerating halo CMEs with
such edges might be shock candidates. A worthwhile shock
indication might be such a halo edge crossing a distorted
streamer. An important complement of morphological suggestions is
kinematics of an expected shock wave and its correspondence to the
drift of a type~II burst. Paper~III also address changes in shape
of a shock front occurring in its propagation in the corona.

\subsection{Presumable Scenario}

Our observations and considerations suggest the following
presumable scenario of a flare-related eruptive event. An eruption
occurs due to a rapid development of an instability in a magnetic
structure. An abruptly accelerating eruption destroys a
pre-existing magnetic configuration, thus causing a flare, and
produces an MHD disturbance as an impulsive piston. The
disturbance appears at a height of $\sim 50$ Mm during the rise
phase of an HXR/microwave burst, leaves the piston, rapidly
steepens into a shock, and then freely propagates like a blast
wave. Its lower trail might be observed as a decelerating Moreton
wave as well as an EUV wave, and the wave dome is sometimes
observed to expand above the limb.

The motion of the shock front shows up in radio spectra as a
drifting continuum and, when the shock front reaches the current
sheet of a coronal streamer, as a type II burst. For the
fundamental emission this usually occurs at $\sim 100$ MHz ($r
\sim 1.5R_{\odot}$). Metric type II bursts are expected to cease
due to damping of shock waves that typically occurs above 20 MHz
($r < 3R_{\odot}$). Revival of a shock is possible at a few
$R_{\odot}$ due to decreasing Alfv{\'e}n speed, and
decametric/hectometric type II emission can appear. The complex
piston-blast-piston transformations of shock waves traveling in
the corona with significantly varying parameters and possible
coupling of multiple shocks imply well-known disagreement between
metric and interplanetary type II events (\textit{e.g.},
\opencite{Cane2005}).

Expanding shock fronts can form envelopes of CMEs. Measurements in
the CME Catalog referring to a fastest feature might be related to
shock waves for fast decelerating CMEs, especially halos. Since a
shock wave decelerates, a trailing mass must eventually approach
its front. The shock becomes a piston-driven one presumably at
distances $r > 20R_{\odot}$. The aerodynamic drag becomes
important. This picture is consistent with results of several
cited papers. The story of shock waves associated with
flare-related CMEs appears to be more complex than often assumed,
in fact combining different scenarios.

\section{Concluding remarks}
\label{S-summary}

Our seemingly simplified approach has resulted in surprisingly
fine reconciliation of EUV waves, Moreton waves, metric type II
bursts, and leading edges of CMEs. The first consequence is that
independent of the quality of our approximation, all these
phenomena are really manifestations of a common agent,
\textit{i.e.}, a traveling coronal shock wave excited by an
eruption. Second, our approach indeed provides a promising
instrument for analyses slow-drifting bursts and their comparison
with other eruption-related phenomena. The power-law approximation
turns out to work well beyond conditions, for which it was
derived. Our results clarify relations between flares, traveling
coronal shocks, CMEs, associated wave-like manifestations, type II
bursts, and provide a common quantitative description for some of
these phenomena. Our important by-product is an indication of the
leading role of eruptions with respect to flares, \textit{i.e.},
that the acceleration of an eruption occurs almost independently
of the flare reconnection rate.

Our approach and analysis needs elaboration and continuation. A
number of issues to be addressed remains. Low heights, at which
type II emission sometimes appears, indicate that our
consideration of its generation in the current sheet of a coronal
streamer needs elaboration. Data sets similar to those analyzed in
our paper should be compared with imaging observations in the
metric range. The analysis of decimetric to metric drifting bursts
should be extended to longer radio waves in conjunction with
coronagraphic observations. Despite success of our self-similar
shock approximation, a more realistic weak shock approximation
should be considered. The last issue is a subject of our paper II.

\begin{acks}

We thank A. Warmuth for data, which he made available to us, and
M.~Eselevich, E.~Ivanov, E.~Schmahl, M.~Temmer, V.~Eselevich,
A.~Altyntsev, G.~Rudenko, L.~Kashapova, V.~Fainshtein,
N.~Prestage, S.~Pohjolainen, S.~White, A.~Zhukov, and J.
Magdaleni{\' c} for fruitful discussions and cooperation. We
gratefully remember Mukul Kundu who inspired a significant part of
our study. We thank an anonymous referee for valuable remarks.

We thank the teams operating all instruments whose data are used
in our study for their efforts and open data policies: the ESA \&
NASA EIT, LASCO, and MDI instruments on SOHO; TRACE and
STEREO/SECCHI telescopes; the Mauna Loa Solar Observatory; the
NOAA/SEC GOES satellites; the NICT HIRAS (Japan), the IPS Radio
and Space Services Learmonth Observatory (Australia), and the USAF
RSTN radio telescopes. We appreciatively use the CME catalog
generated and maintained at the CDAW Data Center by NASA and the
Catholic University of America in cooperation with the Naval
Research Laboratory. This research was supported by the Russian
Foundation of Basic Research under grant 09-02-00115.

\end{acks}

\end{article}

\end{document}